\documentclass[12pt]{article}
\usepackage{graphicx}
\begin{document}

\begin{center}

{\Large \bf Entangled Harmonic Oscillators and Space-time Entanglement}

\vspace{3mm}

Sibel Ba{\c s}kal \footnote{electronic address:
baskal@newton.physics.metu.edu.tr}\\
Department of Physics, Middle East Technical University,
06531 Ankara, Turkey

\vspace{3mm}
Young S. Kim\footnote{electronic address: yskim@physics.umd.edu}\\
Center for Fundamental Physics, University of Maryland,\\
College Park, Maryland 20742, U.S.A.

\vspace{3mm}
Marilyn E. Noz\footnote{electronic address: mairlyne.noz@gmail.com}\\
Department of Radiology,\\
New York University New York, New York 10016, U.S.A.

\end{center}

\vspace{2mm}

\begin{abstract}

The mathematical basis for the Gaussian entanglement is discussed in
detail, as well as its implications in the internal space-time structure of
relativistic extended particles.  It is shown that the Gaussian entanglement
shares the same set of mathematical formulas with the harmonic oscillator in
the Lorentz-covariant world.  It is thus possible to transfer the concept of
entanglement to the Lorentz-covariant picture of the bound state which requires
both space and time separations between two constituent particles.
These space and time variables become entangled as the bound state moves
with a relativistic speed.  It is shown also that our inability to measure
the time-separation variable leads to an entanglement entropy together with
a rise in the temperature of the bound state.
As was noted by Paul A. M. Dirac in 1963, the system of two oscillators contains
the symmetries of $O(3,2)$ de Sitter group containing two $O(3,1)$ Lorentz groups
as its subgroups.  Dirac noted also that the system contains the symmetry of
the $Sp(4)$ group which serves as the basic language for two-mode squeezed
states.  Since the $Sp(4)$ symmetry contains both rotations and squeezes,  one
interesting case is the combination of rotation and squeeze resulting in a shear.
While the current literature is mostly on the entanglement based on squeeze
along the normal coordinates, the shear transformation is an interesting future
possibility.  The mathematical issues on this problem are clarified.

\end{abstract}

\newpage
\section{Introduction}

Entanglement problems deal with fundamental issues in physics. Among them,
the Gaussian entanglement is of current interest not only in
quantum optics~\cite{gied03,braun05,kn05job,ge15} but also in other dynamical
systems~\cite{kn05job,ging02,dodd04,ferra05,adesso07}.
The underlying mathematical language for this form of entanglement is that
of harmonic oscillators.  In this paper, we present first the mathematical
tools which are and may be useful in this branch of physics.
\par
The entangled Gaussian state is based on the formula
\begin{equation} \label{seri00}
\frac{1}{\cosh\eta}\sum_{k} (\tanh{\eta})^k \chi_{k}(x) \chi_{k}(y) ,
\end{equation}
where $\chi_{n}(x)$ is the $n^{th}$ excited-state oscillator wave function.
\par

In Chapter 16 of their book~\cite{walls08}, Walls and Milburn discussed in
detail the role of this formula in the theory of quantum information.
Earlier, this formula played the pivotal role for Yuen to formulate his
two-photon coherent states or two-mode squeezed states~\cite{yuen76}.
The same formula was used by Yurke and Patasek in 1987~\cite{yurke87} and
by Ekert and Knight~\cite{ekert89} for the two-mode squeezed state where
one of the photons is not observed.  The effect of entanglement is to
be seen from the beam splitter experiments~\cite{paris99,mskim02}.
\par

In this paper, we point out first that the series of Eq.(\ref{seri00}) can
also be written  as a squeezed Gaussian form
\begin{equation}\label{i01}
\frac{1}{\sqrt{\pi}}
              \exp{\left\{-\frac{1}{4}\left[e^{-2\eta}(x + y)^2 +
               e^{2\eta}(x - y)^{2}\right]\right\}} ,
\end{equation}
which becomes
\begin{equation}\label{i02}
\frac{1}{\sqrt{\pi}}
              \exp{\left\{-\frac{1}{2}\left(x^2 + y^2\right)\right\}} ,
\end{equation}
when $\eta = 0$.
\par
We can obtain the squeezed form of Eq.(\ref{i01})
by replacing $x$ and $y$ by $x'$ and $y'$ respectively, where
\begin{equation}\label{sq01}
\pmatrix{x' \cr y'} = \pmatrix{\cosh\eta
& -\sinh\eta \cr -\sinh\eta & \cosh\eta}
\pmatrix{x \cr y} .
\end{equation}
If $x$ and $y$ are replaced by $z$ and $t$, Eq.(\ref{sq01}) becomes the
formula for the Lorentz boost along the $z$ direction.  Indeed, the Lorentz
boost is a squeeze transformation~\cite{kn05job,hkn90}.

 \par
\begin{figure}
\centerline{\includegraphics[scale=3.3]{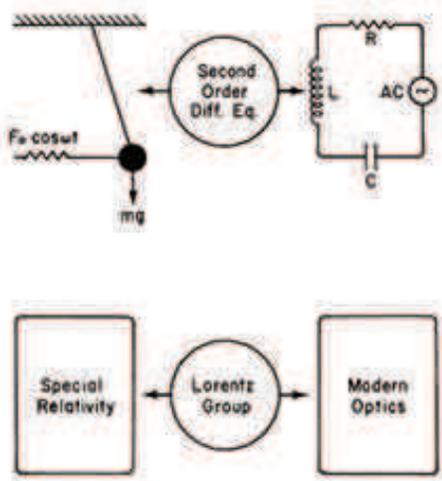}}
\caption{One mathematics for two branches of physics.  Let us look at
Eq.(\ref{seri00}) and Eq.(\ref{i01}) applicable to quantum optics and
special relativity respectively.  They are the same formula from the Lorentz
group with different variables as in the case of the LCR circuit and the mechanical
oscillator sharing the same second-order differential equation.}\label{resonance}
\end{figure}

The squeezed Gaussian form of Eq.(\ref{i01}) plays the key role in
studying boosted bound states in the Lorentz-covariant
world~\cite{dir27,dir45,dir49,yuka53,knp86}, where $z$ and $t$ are the space
and time separations between two constituent particles.  Since the mathematics
of this physical system is the same as the series given in Eq.(\ref{seri00}),
the physical concept of entanglement can be transferred to the Lorentz-covariant
bound state, as illustrated in Fig.~\ref{resonance}.

\par
We can approach this problem from the system of two harmonic oscillators.  In 1963,
Paul A. M. Dirac studied the symmetry of this two-oscillator system and discussed
all possible transformations applicable to this oscillator~\cite{dir63}.
He concluded that there are ten possible generators of transformations satisfying
a closed set of commutation relations.  He then noted that this closed set
corresponds to the Lie algebra of the $O(3,2)$ de Sitter group which is the Lorentz
group applicable to three space-like and two time-like dimensions.  This $O(3,2)$
group has two $O(3,1)$ Lorentz groups as its subgroups.
\par
We note that the Lorentz group is the language of special relativity, while the
harmonic oscillator is one of the major tools for interpreting bound states.
Therefore, Dirac's two-oscillator system can serve as a mathematical framework
for understanding quantum bound systems in the Lorentz-covariant world.

\par
Within this formalism, the series given in Eq.(\ref{seri00}) can be produced from
the ten-generator Dirac system.  In discussing the oscillator system, the
standard procedure is to use the normal coordinates defined as
\begin{equation}\label{norm00}
u = \frac{x + y}{\sqrt{2}}, \quad\mbox{and}\quad v = \frac{x - y}{\sqrt{2}}.
\end{equation}
In terms of these variables, the transformation given in Eq.(\ref{sq01}) takes
the form
\begin{equation} \label{sq02}
\pmatrix{u' \cr v'} = \pmatrix{e^{-\eta} & 0 \cr
   0 & e^{\eta}}
  \pmatrix{u \cr v} ,
\end{equation}
where this is a squeeze transformation along the normal coordinates.  While
the normal-coordinate transformation is a standard procedure, it is interesting
to note that it also serves as a Lorentz boost~\cite{dir49}.

\par
With these preparations, we shall study in Sec.~\ref{2dim}, the system of two
oscillators and  coordinate transformations of current interest.
It is pointed out in Sec.~\ref{dirac63} that there are ten different generators
for transformations, including those discussed in Sec.~\ref{2dim}.  It
is noted that Dirac derived ten generators of transformations applicable to
these oscillators, and they satisfy the closed set of commutation relations
which is the same as  the Lie algebra of the $O(3,2)$ de Sitter group containing
two Lorentz groups among its subgroups. In Sec.~\ref{wigf}, Dirac's ten-generator
symmetry is studied in the Wigner phase-space picture, and it is shown that
Dirac's symmetry contains both canonical and Lorentz transformations.
\par
While the Gaussian entanglement starts from the oscillator wave function in
its ground state, we study in Sec.~\ref{excited} the entanglements of excited
oscillator states.  We give a detailed explanation of how the series
of Eq.(\ref{seri00}) can be derived from the squeezed Gaussian function
of Eq.(\ref{i01}).

In Sec.~\ref{shear}, we study in detail how the sheared state can be derived
from a squeezed state.  It appears to be a rotated squeezed state, but this is
not the case. In Sec.~\ref{restof}, we study what happens when one of the two
entangled variables is not observed within the framework of Feynman's rest of
the universe~\cite{fey72,hkn99ajp}.

\par
In Sec.~\ref{spt}, we note that most of the mathematical formulas in this paper
have been used earlier for understanding relativistic extended particles in the
Lorentz-covariant harmonic oscillator
formalism~\cite{kn73,knp86,kno79jmp,kno79ajp,kiwi90pl,kn11symm}.  These formulas
allow us to transport the concept of
entanglement from the current problem of physics to  quantum bound states in
the Lorentz-covariant world.   The time separation between the constituent particles
is not observable, and is not known in the present form of quantum mechanics.
However, this variable gives its effect in the real world by entangling itself
with the longitudinal variable.

\section{Two-dimensional Harmonic Oscillators}\label{2dim}

The Gaussian form
\begin{equation}\label{wf01}
  \left[\frac{1}{\sqrt{\pi}}\right]^{1/4}\exp{\left(-\frac{x^{2}}{2}\right)}
\end{equation}
is used for many branches of science.  For instance, we can construct
this function by throwing dice.

\par
In physics, this is the wave function for the one-dimensional harmonic
oscillator in the ground state.  This function is also used for the
vacuum state in quantum field theory, as well as the zero-photon state
in quantum optics.  For excited oscillator states, the wave function takes
the form
\begin{equation}\label{wf02}
     \chi_{n}(x) = \left[\frac{1}{\sqrt{\pi}2^n n!}\right]^{1/2}
              H_{n}(x) \exp{\left(\frac{-x^{2}}{2}\right)} ,
\end{equation}
where $H_{n}(x)$ is the Hermite polynomial of the $n^{th}$ degree.
The properties of this wave function are well known, and it
becomes the Gaussian form of Eq.(\ref{wf01}) when $n = 0$.
\par
\begin{figure}
\centerline{\includegraphics[scale=3.0]{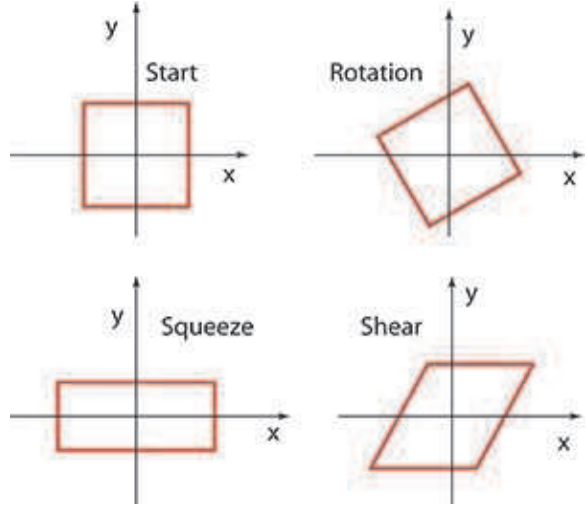}}
\caption{Transformations in the two-dimensional space.  The object
can be rotated, squeezed, or sheared.   In all three cases, the area
remains invariant. }\label{ctran11}
\end{figure}

We can now consider the two-dimensional space with the orthogonal coordinate
variables $x$ and $y$, and the same wave function with the $y$ variable:
\begin{equation}\label{wf05}
     \chi_{m}(y) = \left[\frac{1}{\sqrt{\pi}2^m m!}\right]^{1/2}
              H_{m}(y) \exp{\left(\frac{-y^{2}}{2}\right)} ,
\end{equation}
and construct the function
\begin{equation} \label{wf07}
\psi^{n,m}(x,y) =\left[\chi_{n}(x)\right]\left[\chi_{m}(y)\right].
\end{equation}
This form is clearly separable in the $x$ and $y$ variables.  If $n$ and $m$ are
zero,  the wave function becomes
\begin{equation}\label{gau00}
 \psi^{0,0}(x,y) = \frac{1}{\sqrt{\pi}}
              \exp{\left\{-\frac{1}{2}\left(x^2 + y^2\right)\right\}}.
\end{equation}
Under the coordinate rotation
\begin{equation}\label{trans11}
 \pmatrix{x' \cr y'} = \pmatrix{\cos\theta & -\sin\theta \cr
      \sin\theta & \cos\theta}  \pmatrix{x \cr y}
\end{equation}
this function remains separable.  This rotation is illustrated in
Fig.~\ref{ctran11}.  This is a transformation very familiar to us.
\par
We can next consider  the scale transformation of the form
\begin{equation}\label{sq03}
 \pmatrix{x' \cr y'} = \pmatrix{e^{\eta} & 0 \cr
      0 & e^{-\eta}} \pmatrix{x \cr y} .
\end{equation}
This scale transformation is also illustrated in Fig.~\ref{ctran11}.
This area-preserving transformation is known as the squeeze.  Under
this transformation, the Gaussian function is still separable.
\par
If the direction of the squeeze is rotated by $45^o$, the transformation
becomes the diagonal transformation of Eq.(\ref{sq02}).  Indeed, this is a
squeeze in the normal coordinate system.   This form of squeeze is most
commonly used for squeezed states of light as well as the subject of
entanglements.   It is important to note that, in terms of the $x$ and $y,$
variables, this transformation can be written as Eq.(\ref{sq01})~\cite{dir49}.
In 1905, Einstein used this form of squeeze transformation for the longitudinal
and time-like variables.  This is known as the Lorentz boost.

\par
In addition, we can consider the transformation of the form
\begin{equation}
\pmatrix{x' \cr y'} =
\pmatrix{ 1 & 2\alpha \cr 0 & 1}
\pmatrix{x \cr y} .
\end{equation}
This transformation shears the system as is shown in Fig.~\ref{ctran11}.

\par
After the squeeze or shear transformation, the wave function of Eq.(\ref{wf07})
becomes non-separable, but it can still be written as a series expansion in terms
of the oscillator wave functions. It can take the form
\begin{equation}\label{seri22}
\psi(x, y) = \sum_{n,m} A_{n,m} \chi_{n}(x) \chi_{m}(y) ,
\end{equation}
with
$$
\sum_{n,m} |A_{n,m}|^2 = 1 ,
 $$
if $\psi(x,y)$ is normalized, as was the case for the Gaussian function
of Eq.(\ref{gau00}).

\par

\subsection{Squeezed Gaussian Function}
\begin{figure}
\centerline{\includegraphics[scale=2.5]{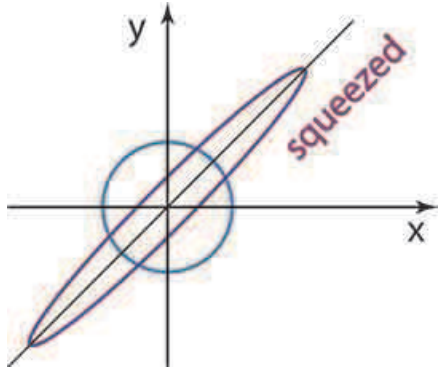}}
\caption{Squeeze along the $45^o$ direction,  discussed most
frequently in the literature.}\label{sqz00}.
\end{figure}

Under the squeeze along the normal coordinate, the Gaussian form of Eq.(\ref{gau00})
becomes
\begin{equation}\label{gau03}
 \psi_{\eta}(x,y) = \frac{1}{\sqrt{\pi}}
              \exp{\left\{-\frac{1}{4}\left[e^{-2\eta}(x + y)^2 +
               e^{2\eta}(x - y)^{2}\right]\right\}} ,
\end{equation}
which was given in Eq.(\ref{i01}). This function is not separable in the $x$ and
$y$ variables.  These variables are now entangled.  We obtain this form by replacing,
in the Gaussian function of Eq.(\ref{gau00}), the $x$ and $y$ variables by $x'$ and
$y'$ respectively, where
\begin{equation}\label{trans22}
  x' = (\cosh\eta) x -  (\sinh\eta) y, \quad\mbox{and}\quad
  y'=  ( \cosh\eta) y  - (\sinh\eta) x .
\end{equation}
This form of squeeze is illustrated in Fig.~\ref{sqz00}, and the expansion
of this squeezed Gaussian function becomes the series given in
Eq.(\ref{seri00})~\cite{knp86,kno79ajp}.  This aspect will be discussed in
detail in Sec.~\ref{excited}.

\par
\par
In 1976~\cite{yuen76}, Yuen discussed two-photon coherent states, often
called squeezed states of light.  This series expansion served as the
starting point for two-mode squeezed states.
More recently, in 2003, Giedke {\it et al.}~\cite{gied03} used this formula
to formulate the concept of the Gaussian entanglement.
\par

There is another way to derive the series.  For the harmonic oscillator
wave functions, there are step-down and step-up operators~\cite{dir45}.
These are defined as
\begin{equation}
a = \frac{1}{\sqrt{2}}\left(x + \frac{\partial}{\partial x} \right),
              \quad\mbox{and}\quad
  a^{\dag} = \frac{1}{\sqrt{2}} \left(x - \frac{\partial}{\partial x}\right) .
\end{equation}
If they are applied to the oscillator  wave function, we have
\begin{equation}
 a~\chi_n(x) = \sqrt{n}~\chi_{n-1}(x),  \quad\mbox{and}\quad
 a^{\dag}~\chi_n(x) = \sqrt{n+1}~ \chi_{n + 1}(x) .
\end{equation}
Likewise, we can introduce $b$ and $b^{\dag} $ operators
applicable to $\chi_{n}(y)$:
\begin{equation}
b = \frac{1}{\sqrt{2}}\left( y + \frac{\partial}{\partial y}\right),
            \quad\mbox{and}\quad
b^{\dag} = \frac{1}{\sqrt{2}} \left(y - \frac{\partial}{\partial y}\right) .
\end{equation}
Thus
\begin{eqnarray}
&{}& \left(a^{\dag}\right)^{n} \chi^{0}(x) = \sqrt{n!}~ \chi_{n}(x), \nonumber \\[1ex].
&{}& \left(b^{\dag}\right)^{n} \chi^{0}(y) = \sqrt{n!} \chi_{n}(y) ,
\end{eqnarray}
and
\begin{equation}
 a~\chi_{0}(x) = b~\chi_{0}(y) = 0 .
\end{equation}
In terms of these variables, the transformation leading the Gaussian function
of Eq.(\ref{gau00}) to its squeezed form of Eq.(\ref{gau03}) can be
written as
\begin{equation}\label{exp01}
\exp{\left\{\frac{\eta}{2}\left( a^{\dag}b^{\dag} - a~b \right)\right\}} ,
\end{equation}
which can also be written as
\begin{equation}\label{exp02}
\exp{\left\{-\eta\left(x\frac{\partial}{\partial y} +
         y\frac{\partial}{\partial x}\right)\right\}} .
\end{equation}

\par
Next, we can consider the exponential form
\begin{equation}\label{exp05}
  \exp{\left\{(\tanh\eta) a^{\dag} b^{\dag}) \right\}} ,
\end{equation}
which can be expanded as
\begin{equation}\label{seri11}
    \sum_{n} \frac{1}{n!} (\tanh\eta)^{n} \left(a^{\dag} b^{\dag}\right)^{n}.
\end{equation}
If this operator is applied to the ground state of Eq.(\ref{gau00}), the
result is
\begin{equation}\label{seri12}
    \sum_{n} (\tanh\eta)^{n} \chi_{n}(x) \chi_{n}(y) .
\end{equation}
This form is not normalized, while the series of Eq.(\ref{seri00}) is.
What is the origin of this difference?
\par
There is a similar problem with the one-photon coherent state~\cite{klaud73,sal07}.
There, the series comes from the expansion of the exponential form
\begin{equation}\label{exp22}
\exp{\left\{\alpha a^{\dag}\right\}} ,
\end{equation}
which can be expanded to
\begin{equation}
\sum_{n} \frac{1}{n!} \alpha^{n} \left(a^{\dag}\right)^{n} .
\end{equation}
However, this operator is not unitary.  In order to make this series unitary,
we consider the exponential form
\begin{equation}
\exp{\left(\alpha a^{\dag} - \alpha^{*} a \right)}
\end{equation}
which is unitary.  This expression  can then be written as
\begin{equation}
e^{-\alpha\alpha^*/2}
\left[ \exp{ \left(\alpha a^{\dag}\right) }  \right ]
\left[\exp{\left(\alpha^* a \right)} \right],
\end{equation}
according to the Baker-Campbell-Hausdorff (BCH) relation~\cite{miller72,hall03}.
If this is applied to the ground state, the last bracket can be
dropped, and the result is
\begin{equation}
e^{-\alpha\alpha^*/2} \exp{\left[\alpha a^{\dag} \right]} ,
\end{equation}
which is the unitary operator with the normalization constant
$$
e^{-\alpha\alpha^*/2}.
$$
\par
Likewise, we can conclude that the series of Eq.(\ref{seri12}) is different
from that of Eq.(\ref{seri00}) due to the difference between the unitary operator
of Eq.(\ref{exp01}) and the non-unitary operator of Eq.(\ref{exp05}).  It may be
possible to derive the normalization factor using the BCH formula, but it seems
to be intractable at this time.  The best way to resolve this problem is to present
the exact calculation of the unitary operator leading to the normalized series of
Eq.(\ref{gau00}).  We shall return to this problem in Sec.~\ref{excited}, where
squeezed excited states are studied.

\subsection{Sheared Gaussian Function}
In addition, there is a transformation called ``shear,'' where only one of
the two
coordinates is translated as shown in Fig.~\ref{ctran11}.
This transformation takes the form
\begin{equation}\label{shr01}
\pmatrix{x' \cr y'} = \pmatrix{1 & 2\alpha \cr  0 & 1} \pmatrix{x \cr y} ,
\end{equation}
which leads to
\begin{equation}\label{shr02}
 \pmatrix{x' \cr y'} =  \pmatrix{x + 2\alpha y \cr y} .
 \end{equation}
This shear is one of the basic transformations in engineering sciences.
In physics, this transformation plays the key role in understanding
the internal space-time symmetry of massless
particles~\cite{wig39,wein64,kiwi90jmp}.  This matrix plays the
pivotal role during the transition from the oscillator mode to the
damping mode in classical damped harmonic oscillators~\cite{bkn14,bkn15}.
\par
Under this transformation, the Gaussian form becomes
\begin{equation}\label{gau33}
 \psi_{shr}(x,y) = \frac{1}{\sqrt{\pi}}
 \exp{\left\{-\frac{1}{2}\left[(x - 2\alpha y)^2 + y^2\right]\right\}}.
\end{equation}
It is possible to expand this into a series of the form of
Eq.(\ref{seri22})~\cite{kimyeh92}.
\par
The transformation applicable to the Gaussian form of Eq.(\ref{gau00})
is
\begin{equation}\label{shr05}
\exp{\left(-2\alpha y \frac{\partial}{\partial x} \right)}  ,
\end{equation}
and the generator is
\begin{equation}\label{shr07}
   - i y \frac{\partial}{\partial x}  .
\end{equation}
It is of interest to see where this generator stands among the ten
generators of Dirac.
\par
However, the most pressing problem is whether the sheared Gaussian form
can be regarded as a rotated squeezed state.  The basic mathematical issue
is that the shear matrix of Eq.(\ref{shr01}) is triangular and cannot be
diagonalized.  Therefore, it cannot be a squeezed state. Yet, the Gaussian
form of Eq.(\ref{gau33}) appears to be a rotated squeezed state, while
not along the normal coordinates. We shall look at this problem in detail
in Sec.~\ref{shear}.

\section{Dirac's Entangled Oscillators}\label{dirac63}

Paul A. M. Dirac devoted much of his life-long efforts to the task of making
quantum mechanics compatible with special relativity.  Harmonic oscillators
serve as an instrument for illustrating quantum mechanics, while special
relativity is the physics of the Lorentz group.  Thus, Dirac attempted to
construct a representation of the Lorentz group using harmonic oscillator
wave functions~\cite{dir45,dir63}.
\par
In his 1963 paper~\cite{dir63}, Dirac started from the two-dimensional oscillator
whose wave function takes the Gaussian form given in Eq.(\ref{gau00}).  He then
considered unitary transformations applicable to this ground-state wave function.
He noted that they can be generated by the following ten Hermitian operators
\begin{eqnarray}\label{dir10}
&{}& L_{1} = {1\over 2}\left(a^{\dag} b + b^{\dag} a\right) ,\qquad
L_{2} = {1\over 2i}\left(a^{\dag} b - b^{\dag} a\right) ,  \nonumber \\[3mm]
&{}& L_{3} = {1\over 2}\left(a^{\dag} a - b^{\dag} b \right) , \qquad
S_{3} = {1\over 2}\left(a^{\dag}a + bb^{\dag}\right) ,   \nonumber \\[3mm]
&{}& K_{1} = -{1\over 4}\left(a^{\dag}a^{\dag} + aa - b^{\dag}b^{\dag} - bb\right) , \nonumber \\[3mm]
&{}& K_{2} = {i\over 4}\left(a^{\dag}a^{\dag} - aa + b^{\dag}b^{\dag} - bb\right) ,   \nonumber \\[3mm]
&{}& K_{3} = {1\over 2}\left(a^{\dag}b^{\dag} + ab\right) ,   \nonumber \\[3mm]
&{}& Q_{1} = -{i\over 4}\left(a^{\dag}a^{\dag} - aa - b^{\dag}b^{\dag} + bb \right) ,   \nonumber \\[3mm]
&{}& Q_{2} = -{1\over 4}\left(a^{\dag}a^{\dag} + aa + b^{\dag}b^{\dag} + bb \right) ,  \nonumber \\[3mm]
&{}& Q_{3} = {i\over 2}\left(a^{\dag}b^{\dag} - ab \right) .
\end{eqnarray}
He then noted that these operators satisfy the following set of commutation
relations.
\begin{eqnarray}\label{dir12}
&{}& [L_{i}, L_{j}] = i\epsilon _{ijk} L_{k} ,\qquad
[L_{i}, K_{j}] = i\epsilon _{ijk} K_{k} , \qquad
[L_{i}, Q_{j}] = i\epsilon _{ijk} Q_{k} , \nonumber\\[1ex]
&{}&
[K_{i}, K_{j}] = [Q_{i}, Q_{j}] = -i\epsilon _{ijk} L_{k} ,  \qquad
[L_{i}, S_{3}] = 0 ,  \nonumber\\[1ex]
&{}&
[K_{i}, Q_{j}] = -i\delta _{ij} S_{3} , \qquad
[K_{i}, S_{3}] =  -iQ_{i} , \qquad [Q_{i}, S_{3}] = iK_{i} .
\end{eqnarray}

\par
Dirac then determined that these commutation relations constitute the
Lie algebra for the $O(3,2)$ de Sitter group with ten generators. This de Sitter
group is the Lorentz group applicable to three space coordinates and two time
coordinates.  Let us use the notation $(x, y, z, t, s)$, with $(x, y, z)$ as space
coordinates and $(t, s)$ as two time coordinates.  Then the rotation around the
$z$ axis is generated by
\begin{equation}\label{dir15}
L_{3} = \pmatrix{0 & -i & 0 & 0 & 0 \cr i & 0 & 0 & 0 & 0 \cr
  0 & 0 & 0 & 0 & 0 \cr 0 & 0 & 0 & 0 & 0 \cr 0 & 0 & 0 & 0 & 0 } .
\end{equation}
The generators $L_1$ and $L_2$ can be also be constructed.  The $K_3$
and $Q_3$ generators will take the form
\begin{equation}
K_{3} = \pmatrix{0 & 0 & 0 & 0 & 0 \cr 0 & 0 & 0 & 0 & 0 \cr
  0 & 0 & 0 & i & 0 \cr 0 & 0 & i & 0 & 0 \cr 0 & 0 & 0 & 0 & 0 } , \qquad
Q_{3} =\pmatrix{0 & 0 & 0 & 0 & 0 \cr 0 & 0 & 0 & 0 & 0 \cr
               0 & 0 & 0 & 0 & i \cr 0 & 0 & 0 & 0 & 0 \cr
               0 & 0 & i & 0 & 0 } .
\end{equation}
From these two matrices, the generators $K_1, K_2, Q_1, Q_2$ can be
constructed.  The generator $S_3$ can be written as
\begin{equation} \label{dir18}
S_{3} = \pmatrix{0 & 0 & 0 & 0 & 0 \cr 0 & 0 & 0 & 0 & 0 \cr
  0 & 0 & 0 & 0 & 0 \cr 0 & 0 & 0 & 0 & -i \cr 0 & 0 & 0 & i &  0 } .
\end{equation}
The last five-by-five matrix generates rotations in the two-dimensional space of
$(t, s)$.   If we introduce these two time variables, the $O(3,2)$ group leads
to two coupled Lorentz groups.  The particle mass is invariant under Lorentz
transformations.  Thus, one Lorentz group cannot change the particle mass.
However, with two coupled Lorentz groups we can describe the world with variable
masses such as the neutrino oscillations.

\par
In Sec.~\ref{2dim}, we used the operators $Q_{3}$ and $K_{3}$ as the
generators for the squeezed Gaussian function.  For the unitary transformation of
Eq.(\ref{exp01}), we used
\begin{equation}\label{exp11}
\exp{\left(-i\eta Q_{3}\right)} .
\end{equation}
However, the exponential form of Eq.(\ref{exp05}) can be written as
\begin{equation}\label{exp12}
\exp{\left\{-i(\tanh\eta) \left(Q_{3} + iK_{3}\right)\right\}} ,
\end{equation}
which is not unitary, as was seen before.

\par
From the space-time point of view, both $K_{3}$ and $Q_{3}$ generate
Lorentz boosts along the $z$ direction, with the time variables
$t$ and $s$ respectively.  The fact that the squeeze and Lorentz
transformations share the same mathematical formula is well known.
However, the non-unitary operator $iK_{3}$ does not seem to have
a space-time interpretation.

\par

As for the sheared state, the generator can be written as
\begin{equation}
     Q_{3} - L_{2} ,
\end{equation}
leading to the expression given in Eq.(\ref{shr07}).  This is a Hermitian
operator leading to the unitary transformation of Eq.(\ref{shr05}).

\section{Entangled Oscillators in the Phase-Space Picture}\label{wigf}

Also in his 1963 paper, Dirac states that the Lie algebra of Eq.~(\ref{dir12})
can serve as the four-dimensional symplectic group $Sp(4)$.  This group
allows us to study squeezed or entangled states in terms of the four-dimensional
phase space consisting of two position and two momentum
variables~\cite{hkn90,knp91,kn13}.

\par

\begin{figure}
\centerline{\includegraphics[scale=3.5]{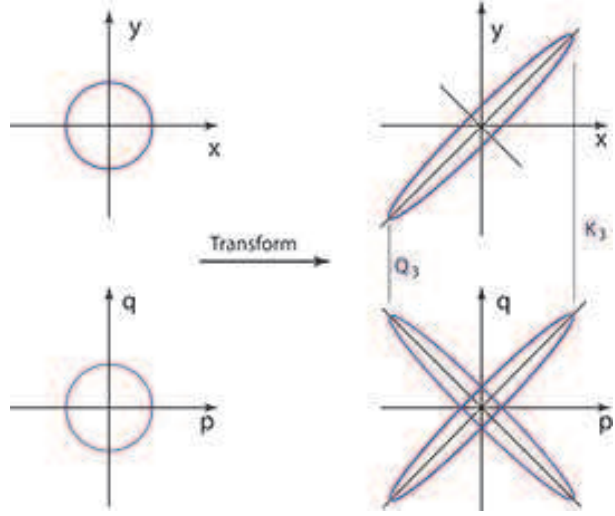}}
\caption{Transformations generated by $Q_{3}$ and $K_{3}$.  As the
parameter $\eta$ becomes larger, both the space and momentum distribution
becomes larger. }\label{ctran33}
\end{figure}

In order to study the $Sp(4)$ contents of the coupled oscillator system,
let us introduce the Wigner function defined as~\cite{wig32}
\begin{eqnarray}\label{wigf01}
\lefteqn{W(x,y; p,q) = \left({1\over \pi} \right)^{2}
\int \exp \left\{- 2i (px' + qy') \right\} } \nonumber\\[3mm]
\mbox{ } & \mbox{ } & \mbox{ }
\times \psi^{*}(x + x', y + y')
\psi (x - x', y - y') dx' dy' .\hspace*{2cm}
\end{eqnarray}
If the wave function $\psi(x,y)$ is the Gaussian form of Eq.(\ref{gau00}),
the Wigner function becomes
\begin{equation}\label{wigf00}
W(x,y:p,q) = \left(\frac{1}{\pi}\right)^{2}
          \exp{\left\{-\left(x^2 + p^2 + y^2 + q^2\right)\right\} } .
\end{equation}
\par
The Wigner function is defined over the four-dimensional phase space of
$(x, p, y, q)$ just as in the case of classical mechanics.  The unitary
transformations generated by the operators of Eq.(\ref{dir10}) are
translated into Wigner transformations~\cite{knp91,hkn95jmp,kn13}.
As in the case of Dirac's oscillators, there are ten corresponding
generators applicable to the Wigner function.  They are
\begin{eqnarray}\label{rotphase}
L_{1} &=& +{i\over 2}\left\{\left(x{\partial \over \partial q} -
q{\partial \over \partial x} \right) +
\left(y{\partial \over \partial p} -
p{\partial \over \partial y} \right)\right\}, \nonumber \\[3mm]
L_{2} &=& -{i\over 2}\left\{\left(x{\partial \over \partial y} -
y{\partial \over \partial x}\right) +
\left(p {\partial \over \partial q} -
q{\partial \over \partial p}\right)\right\} ,\nonumber \\[3mm]
L_{3} &=& +{i\over 2}\left\{\left(x{\partial \over \partial p} -
p{\partial \over \partial x}\right) -
\left(y{\partial \over \partial q} -
q{\partial \over \partial y}\right)\right\} , \nonumber \\[3mm]
S_{3} &=& -{i\over 2}\left\{\left(x{\partial \over \partial p} -
p{\partial \over \partial x}\right) +
\left(y{\partial \over \partial q} -
q{\partial \over \partial y}\right)\right\} ,
\end{eqnarray}
and
\begin{eqnarray}\label{sqphase}
K_{1} &=& -{i\over 2}\left\{\left( x{\partial \over \partial p} +
p{\partial \over \partial x} \right) -
\left(y{\partial \over \partial q} +
q{\partial \over \partial y} \right)\right\}, \nonumber \\[3mm]
K_{2} &=& -{i\over 2}\left\{\left(x{\partial \over \partial x} +
  y{\partial \over \partial y}\right) -
  \left(p{\partial \over \partial p} +
  q{\partial \over \partial q}\right)\right\} , \nonumber \\[3mm]
K_{3} &=& +{i\over 2}\left\{\left(x{\partial \over \partial q} +
  q{\partial \over \partial x}\right) +
  \left(y{\partial \over \partial p} +
  p{\partial \over \partial y}\right)\right\} , \nonumber \\[3mm]
Q_{1} &=& +{i\over 2}\left\{\left(x{\partial \over \partial x} +
  q{\partial \over \partial q}\right) -
  \left(y{\partial \over \partial y} +
  p{\partial \over \partial p}\right)\right\} ,\nonumber \\[3mm]
Q_{2} &=& -{i\over 2}\left\{\left(x{\partial \over \partial p} +
  p{\partial \over \partial x}\right) +
  \left(y{\partial \over \partial q} +
  q{\partial \over \partial y}\right)\right\} ,\nonumber \\[3mm]
Q_{3} &=& -{i\over 2}\left\{\left(y{\partial \over \partial x} +
  x{\partial \over \partial y} \right) -
  \left(q{\partial \over \partial p} + p{\partial \over
  \partial q}\right)\right\} .
\end{eqnarray}
These generators also satisfy the Lie algebra given in Eq.(\ref{dir10})
and Eq.(\ref{dir12}).  Transformations generated by these generators
have been discussed in the literature~\cite{hkn90,hkn95jmp,kn13}.

  \par

As in the case of Sec.~\ref{dirac63}, we are interested in the generators
$Q_{3}$ and  $K_{3}$.  The transformation generated by $Q_{3}$ takes the
form
\begin{equation}
 \left[\exp{\left\{\eta \left(x\frac{\partial}{\partial y} +
     y\frac{\partial}{\partial x}\right)\right\} } \right]
  \left[\exp{\left\{-\eta \left(p\frac{\partial}{\partial q} +
  q\frac{\partial}{\partial p}\right)\right\} }  \right] .
\end{equation}
This exponential form squeezes the Wigner function of Eq.(\ref{wigf00}) in
the $x~y$ space as well as in their corresponding momentum space. However,
in the momentum space, the squeeze is in the opposite direction as illustrated in
Fig.~\ref{ctran33}.  This is what we expect from canonical transformation in
classical mechanics.  Indeed, this corresponds to the unitary transformation
which played the major role in Sec.~\ref{2dim}.

\par
Even though shown insignificant in Sec.~\ref{2dim}, $K_{3}$ had a definite
physical interpretation in Sec.~\ref{dirac63}. The transformation generated by $K_{3}$ takes the
form
\begin{equation}
 \left[\exp{\left\{\eta \left(x\frac{\partial}{\partial q} +
     q\frac{\partial}{\partial x}\right)\right\} } \right]
  \left[\exp{\left\{\eta \left(y\frac{\partial}{\partial p} +
 p\frac{\partial}{\partial y}\right)\right\} }  \right] .
\end{equation}
This performs the squeeze in the $x~q$ and $y~p$ spaces. In this case,
the squeezes have the same sign, and the rate of increase is  the same in
all directions. We can thus have the same picture of squeeze for both $x~y$
and $p~q$ spaces as illustrated in Fig.~\ref{ctran33}.  This parallel
transformation corresponds to the Lorentz squeeze~\cite{knp86,kno79jmp}.
\par
As for the sheared state, the combination
\begin{equation}
Q_{3} -  L_{2}  = -i\left( y\frac{\partial}{\partial x} +
q\frac{\partial}{\partial p} \right) ,
\end{equation}
generates the same shear in the $p~q$ space.

\section{Entangled Excited States}\label{excited}

In Sec.~\ref{2dim}, we discussed the entangled ground state, and noted that the
entangled state of Eq.(\ref{seri00}) is a series expansion of the squeezed
Gaussian function.  In this section, we are interested in what happens
when we squeeze an excited oscillator state starting from
\begin{equation}\label{wf10}
\chi_{n}(x)\chi_{m}(y) .
\end{equation}
In order to entangle this state, we should replace $x$ and $y$ respectively
by $x'$ and $y'$ given in Eq.(\ref{trans22}).

\par
The question is how the oscillator wave function is squeezed after this operation.
Let us note first that the wave function of Eq.(\ref{wf10}) satisfies the
equation
\begin{equation}\label{cov01}
\frac{1}{2}\left\{\left(x^2 - \frac{\partial^2}{\partial x^2}\right)
 - \left(y^2 - \frac{\partial^2}{\partial y^2}\right)\right\}
 \chi_n(x) \chi_m(y) = (n - m) \chi_n(x) \chi_m(y) .
\end{equation}
This equation is invariant under the squeeze transformation of Eq.(\ref{trans22}),
and thus the eigenvalue $(n - m)$ remains invariant.  Unlike the usual two-oscillator
system, the $x$ component and the $y$ component have opposite signs.  This is the
reason why the overall equation is squeeze-invariant~\cite{kno79jmp,kn05job,fkr71}.
\par
We then have to write this squeezed oscillator in the series form of
Eq.(\ref{seri22}).  The most interesting case is of course for $m = n =0 $, which
leads to the Gaussian entangled state given in Eq.(\ref{gau03}).  Another interesting
case is for $m = 0$ while $n$ is allowed to take all integer values.  This
single-excitation system has applications in the covariant oscillator formalism
where no time-like excitations are allowed.   The Gaussian entangled state is a
special case of this single-excited oscillator system.
\par
The most general case is for nonzero integers for both $n$ and $m$.  The calculation
for this case is available in the literature~\cite{knp86,rotbart81}.  Seeing no
immediate physical applications of this case, we shall not reproduce this
calculation in this section.
\par
For the single-excitation system, we write the starting wave function as
\begin{equation}\label{wf12}
     \chi_{n}(x)\chi_{0}(y) = \left[\frac{1}{\pi~2^n n!}\right]^{1/2}
              H_{n}(x) \exp{\left\{-\left(\frac{x^2 + y^2 }{2}\right)\right\}} .
\end{equation}
There are no excitations along the $y$ coordinate.  In order to squeeze this
function, our plan is to replace $x$ and $y$ by $x'$ and $y'$ respectively
and write $\chi_{n}(x')\chi_{0}(y')$ as a series in the form
\begin{equation}
 \chi_{n}(x')\chi_{0}(y') = \sum_{k',k} A_{k',k}(n)  \chi_{k'}(x)\chi_{k}(y) .
\end{equation}
Since $k' - k = n$  or $ k' = n + k$, according to the eigenvalue of
the differential
equation given in Eq.(\ref{cov01}), we write this series as
\begin{equation}
 \chi_{n}(x')\chi_{0}(y') = \sum_{k',k} A_{k}(n)  \chi_{(k + n)}(x)\chi_{k}(y) ,
\end{equation}
with
\begin{equation}
     \sum_{k} |A_{k}(n)|^2 = 1 .
\end{equation}
This coefficient is
\begin{equation} \label{coeff55}
A_{k}(n)  = \int \chi_{k + n}(x)\chi_{k}(y)\chi_{n}(x')\chi_{0}(y')~ dx~dy .
\end{equation}
This calculation was given in the literature in a fragmentary way
in connection with a Lorentz-covariant description of extended particles starting
from Ruiz's 1974 paper~\cite{ruiz74}, subsequently by
Kim {\it et al.} in 1979~\cite{kno79ajp} and by Rotbart in 1981~\cite{rotbart81}.
In view of the recent developments of physics, it seems necessary to give
one coherent calculation of the coefficient of Eq.(\ref{coeff55}).
\par
We are now interested in the squeezed oscillator function
\begin{eqnarray}\label{coef55}
&{}& A_{k}(n) =
\left[\frac{1}{\pi^2 ~2^n n! (k + n)^2 (n + k)! k^2 k!}\right]^{1/2} \nonumber \\[1ex]
&{}& \times \int    H_{n+k}(x) H_{k}(y) H_{n}(x')
\exp{\left\{-\left(\frac{x^2 + y^2 + x'^2 + y'^2 }{2}\right)\right\}}dx dy .
\end{eqnarray}
As was noted by Ruiz~\cite{ruiz74}, the key to the evaluation of this integral
is to introduce the generating function for the Hermite
polynomials~\cite{magnus66,doman16}:
\begin{equation}
G(r,z) = \exp{\left(-r^2 + 2rz\right)} = \sum_{m} \frac{r^m}{m!} H_m(z) ,
\end{equation}
and evaluate the integral
\begin{equation}
I = \int G(r,x) G(s,y) G(r',x')
  \exp{\left\{-\left(\frac{x^2 + y^2 + x'^2 + y'^2 }{2}\right)\right\}}dx dy .
\end{equation}
The integrand becomes one exponential function, and its exponent is quadratic
in $x$ and $y$.  This quadratic form can be diagonalized, and the integral can
be evaluated~\cite{knp86,kno79ajp}.  The result is
\begin{equation}
  I = \left[\frac{\pi}{\cosh\eta}\right] \exp{( 2rs \tanh\eta)}
  \exp{\left(\frac{2rr'}{\cosh\eta}\right) }.
\end{equation}
We can now expand this expression and choose the coefficients of
$r^{n+k}, s^{k}, r'^{n}$  for \\ $H_{(n + k)}(x), H_{n}(y)$, and $H_{n}(z')$
respectively.  The result is
\begin{equation}
 A_{n;k} = \left(\frac{1}{\cosh\eta}\right)^{(n+1)}
     \left[\frac{(n+k)!}{n!k!}\right]^{1/2} (\tanh\eta)^{k}.
\end{equation}
Thus, the series becomes
\begin{equation}\label{seri55}
\chi_{n}(x') \chi_{0}(y') = \left(\frac{1}{\cosh\eta}\right)^{(n+1)}
 \sum_{k} \left[\frac{(n+k)!}{n!k!}\right]^{1/2}
       (\tanh\eta)^{k} \chi_{k+n}(x)\chi_{k}(y) .
\end{equation}
If $n = 0$, it is the squeezed ground state, and this expression becomes
the entangled state of Eq.(\ref{gau03}).

\begin{figure}
\centerline{\includegraphics[scale=4.0]{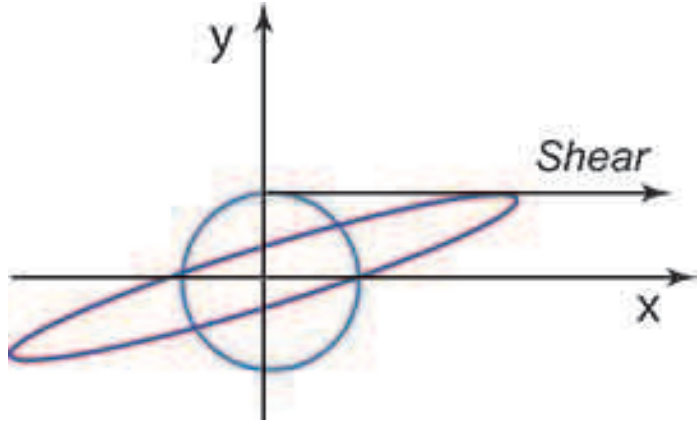}}
\caption{Sear transformation of the Gaussian form given in
Eq.(\ref{gau00}).}\label{shear00}
\end{figure}

\section{E(2)-sheared States}\label{shear}

Let us next consider the effect of shear on the Gaussian form.  From
Fig.~\ref{shear00} and Fig.~\ref{sqz00}, it is clear  that the sheared
state is a rotated squeezed state.
\par
In order to understand this transformation, let us note that the squeeze and
rotation are generated by the two-by-two matrices
\begin{equation}
K = \pmatrix{0 & i \cr i & 0}, \qquad J = \pmatrix{0 & -i \cr i & 0}, \qquad
\end{equation}
which generate the squeeze and rotation matrices of the form
\begin{eqnarray}
&{}& \exp{\left(-i\eta K\right)} =
\pmatrix{\cosh\eta & \sinh\eta \cr \sinh\eta & \cosh\eta },   \nonumber\\[1ex]
&{}& \exp{\left(-i \theta J\right)} = \pmatrix{\cos\theta & -\sin\theta \cr
     \sin\theta &  \cos\theta} ,
\end{eqnarray}
respectively.
We can then consider
\begin{equation}\label{shr05a}
S = K - J = \pmatrix{0 & 2i \cr 0 & 0} .
\end{equation}
This matrix has the property that $S^2 = 0$.  Thus the transformation matrix
becomes
\begin{equation}\label{shr10}
  \exp{(-i\alpha S)} = \pmatrix{1 & 2\alpha \cr 0 & 1} .
\end{equation}
Since $S^2 = 0$, the Taylor expansion truncates, and the transformation matrix
becomes the triangular matrix of Eq.(\ref{shr02}), leading to the transformation
\begin{equation}\label{shr09}
\pmatrix{x \cr y}
\quad\rightarrow\quad \pmatrix{x + 2\alpha y \cr y} .
\end{equation}
\par
The shear generator $S$ of Eq.(\ref{shr05a}) indicates that
the infinitesimal transformation is a rotation followed by a squeeze.
Since both rotation and squeeze are area-preserving transformations,
the shear should also be an area-preserving transformations.

\par
In view of Fig.~\ref{shear00}, we should ask whether the triangular matrix of
Eq.(\ref{shr10}) can be obtained from one squeeze matrix followed by one
rotation matrix.  This is not possible mathematically.  It can however,
 be written as a squeezed rotation matrix of the form

\begin{equation}\label{sqrot01}
\pmatrix{e^{\lambda/2} & 0 \cr 0 &  e^{-\lambda/2}}
\pmatrix{\cos\omega & \sin\omega \cr -\sin\omega & \cos\omega}
\pmatrix{e^{-\lambda/2} & 0 \cr 0 &  e^{\lambda/2}},
\end{equation}
resulting in
\begin{equation}\label{mat63}
\pmatrix{\cos\omega & e^{\lambda}\sin\omega \cr -e^{-\lambda}\sin\omega & \cos\omega}.
\end{equation}
If we let
\begin{equation}
(\sin\omega) = 2\alpha e^{-\lambda}
\end{equation}
Then
\begin{equation} \label{tri05}
\pmatrix{\cos\omega & 2\alpha \cr -2\alpha e^{-2\lambda} & \cos\omega }.
\end{equation}
If $\lambda$ becomes infinite, the angle $\omega$ becomes zero, and  this
matrix becomes the triangular matrix of Eq.(\ref{shr10}).  This is a
singular process where the parameter $\lambda$ goes to infinity.

\par
If this transformation is applied to the Gaussian form of Eq.(\ref{gau00}),
it becomes
\begin{equation}\label{gau05}
 \psi(x,y) = \frac{1}{\sqrt{\pi}}
   \exp{\left\{-\frac{1}{2}\left[(x - 2\alpha y)^2 + y^2\right]\right\}}.
\end{equation}
The question is whether the exponential portion of this expression can be written as
\begin{equation}\label{gau07}
   \exp{\left\{-\frac{1}{2}\left[e^{-2\eta}(x~\cos\theta + y~\sin\theta)^2 +
     e^{2\eta}(x~\sin\theta - y~\cos\theta)^2\right]\right\}} .
\end{equation}
The answer is Yes. This is possible if
\begin{eqnarray}
&{}& \tan(2\theta) = \frac{1}{\alpha}, \nonumber \\[1ex]
&{}& e^{2\eta} = 1 + 2\alpha^2  + 2\alpha \sqrt{\alpha^2 + 1}, \nonumber \\[1ex]
&{}& e^{-2\eta} = 1 + 2\alpha^2  - 2\alpha \sqrt{\alpha^2 + 1}.
\end{eqnarray}

\par
In Eq.(\ref{tri05}), we needed a limiting case of $\lambda$ becoming infinite.
This is necessarily a singular transformation.  On the other hand, the derivation
of the Gaussian form of Eq.(\ref{gau05}) appears to be analytic.  How is it
possible?  In order to achieve the transformation from the Gaussian form of
Eq.(\ref{gau00}) to Eq.(\ref{gau05}), we need the linear transformation
\begin{equation}
       \pmatrix{\cos\theta & -\sin\theta \cr \sin\theta & \cos\theta}
       \pmatrix{e^{\eta} & 0 \cr 0 & e^{-\eta}}.
\end{equation}
If the initial form is invariant under rotations as in the case of the
Gaussian function of Eq.(\ref{gau00}), we can add another rotation
matrix on the right hand side.  We choose that rotation matrix to be
\begin{equation}
       \pmatrix{\cos(\theta - \pi/2) & -\sin(\theta - \pi/2)  \cr
         \sin(\theta - \pi/2) & \cos(\theta - \pi/2)} ,
\end{equation}
write the three matrices as
\begin{equation}
\pmatrix{\cos\theta'  & -\sin\theta' \cr \sin\theta'  & \cos\theta'}
\pmatrix{\cosh\eta  & \sinh\eta  \cr \sinh\eta  & \cosh\eta}
\pmatrix{\cos\theta'  & -\sin\theta' \cr \sin\theta'  & \cos\theta'},
\end{equation}
with
$$
\theta' = \theta - \frac{\pi}{4} .
$$
The multiplication of these three matrices leads to
\begin{equation}\label{mat67}
 \pmatrix{(\cosh\eta)\sin(2\theta) & \sinh\eta + (\cosh\eta)\cos(2\theta) \cr
    \sinh\eta - (\cosh\eta)\cos(2\theta) & (\cosh\eta)\sin(2\theta)} .
\end{equation}
The lower-left element can become zero when $\sinh\eta = \cosh(\eta)\cos(2\theta)$
and consequently this matrix becomes
\begin{equation}
 \pmatrix{1 & 2\sinh\eta \cr 0 & 1} .
\end{equation}
Furthermore, this matrix can be written in the form of a squeezed rotation matrix
given in Eq.(\ref{mat63}), with
\begin{eqnarray}
&{}& \cos\omega = (\cosh\eta)\sin(2\theta),  \nonumber \\[1ex]
&{}& e^{-2\lambda} = \frac{ \cos(2\theta)- \tanh\eta }
                     { \cos(2\theta)+ \tanh\eta } .
\end{eqnarray}
The matrices of the form of Eq.(\ref{mat63}) and Eq.(\ref{mat67}) are known as the
Wigner and Bargmann decompositions respectively~\cite{wig39,barg47,hk88,hkn99,bkn14}.

\section{Feynman's Rest of the Universe}\label{restof}

We need the concept of entanglement in quantum systems of two variables.
The issue is how the measurement of one variable affects the other variable.
The simplest case is what happens to the first variable while no measurements
are taken on the second variable.  This problem has a long history since
von Neumann introduced the concept of density matrix in 1932~\cite{neu32}.
While there are many books and review articles on this subject, Feynman stated
this problem in his own colorful way.  In his book on statistical
mechanics~\cite{fey72}, Feynman makes the following statement about the density
matrix.
\par
{\it When we solve a quantum-mechanical problem, what we
really do is divide the universe into two parts - the system in which we
are interested and the rest of the universe.  We then usually act as if
the system in which we are interested comprised the entire universe.
To motivate the use of density matrices, let us see what happens when we
include the part of the universe outside the system}.

\par

Indeed, Yurke and Potasek~\cite{yurke87}  and also Ekert and Knight~\cite{ekert89}
studied this problem in the two-mode squeezed state using the entanglement
formula given in Eq.(\ref{gau03}).  Later in 1999, Han {\it et al.} studied this
problem with two coupled oscillators where one oscillator is observed while the
other is not and thus is in the rest of the universe as defined by
Feynman~\cite{hkn99ajp}.
\par
Somewhat earlier in 1990~\cite{kiwi90pl}, Kim and Wigner observed that there is
a time separation  wherever there is a space separation in the Lorentz-covariant
world.  The Bohr radius is a space separation.  If the system is Lorentz-boosted,
the time-separation becomes entangled with the space separation.  But, in the
present form of quantum mechanics, this time-separation variable is not measured
and not understood.
\par
This variable  was  mentioned in the paper of Feynman {\it et al.}
in 1971~\cite{fkr71}, but the authors say they would drop this variable because
they do not know what to do with it.  While what Feynman {\it et al.} did
was not quite respectable from the scientific point of view, they made a
contribution by pointing out the existence of the problem.  In 1990, Kim and
Wigner~\cite{kiwi90pl} noted that the time-separation variable belongs to Feynman's
rest of the universe and studied its consequences in the observable world.

\par
In this section, we first reproduce the work of Kim and Wigner using the
$x$ and $y$ variables and then study its consequences.  Let us introduce
the notation $\psi_{\eta}^{n}(x,y)$ for the squeezed oscillator wave function
given in Eq.(\ref{seri55}):
\begin{equation}\label{wf15}
\psi_{\eta}^{n}(x,y) = \chi_{n}(x') \chi_{0}(y'),
\end{equation}
with no excitations along the $y$ direction.  For $\eta = 0$, this expression
becomes $\chi_{n}(x) \chi_{0}(y).$

From this wave function, we can construct the pure-state density matrix as
\begin{equation}\label{eq12}
  \rho_\eta^{n}(x,y;r,s) = \psi_{\eta}^{n}(x,y) \psi_{\eta}^{n} (r,s) ,
\end{equation}
which satisfies the condition $\rho^2  = \rho, $  which means
\begin{equation}
  \rho_\eta^{n}(x,y;r,s) = \int \rho_\eta^{n}(x,y;u,v)
   \rho_\eta^{n}(u,v;r,s) du dv .
\end{equation}

\begin{figure}
\centerline{\includegraphics[scale=4.0]{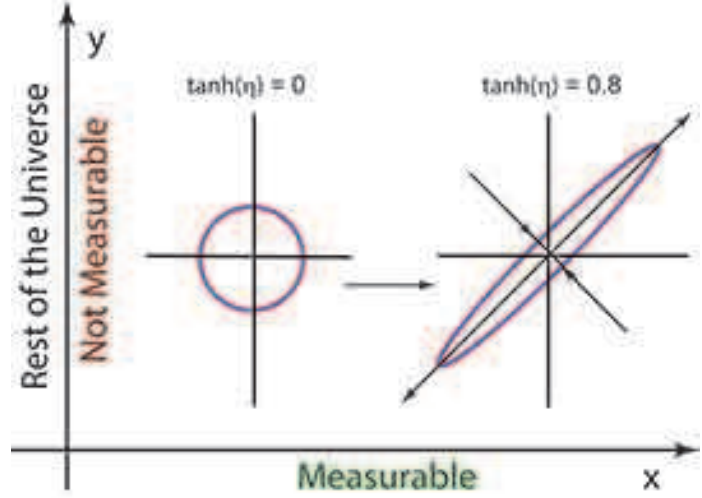}}
\caption{Feynman's rest of the universe.  As the Gaussian function is
squeezed, the $x$ and $y$ variables become entangled.  If the $y$
variable is not measured, it affects the quantum mechanics of the $x$
variable.}\label{restof66}
\end{figure}

As illustrated in Fig.~\ref{restof66}, it is not possible make measurements
on the variable $y.$   We thus have to take the trace of this density matrix
along the $y$ axis, resulting in
\begin{eqnarray}\label{eq14}
&{}& \rho_\eta^{n}(x,r) = \int \psi_{\eta}^{n}(x,y)
                            \psi_{\eta}^{n}(r,y) dy \nonumber \\[2ex]
&{}& \hspace{8mm} = \left(\frac{1}{\cosh\eta}\right)^{2(n + 1)}
     \sum_{k} \frac{(n+k)!}{n!k!}
     (\tanh\eta)^{2k}\chi_{n+k}(x)\chi_{k+n}(r) .
\end{eqnarray}
The trace of this density matrix is one, but the trace of $\rho^2$ is
\begin{eqnarray}
&{}& Tr\left(\rho^2\right) = \int \rho_\eta^{n}(x,r)
                        \rho_\eta^{n}(r,x) dr dx \nonumber \\[2ex]
  &{}& \hspace{10mm} = \left(\frac{1}{\cosh\eta}\right)^{4(n + 1)}
  \sum_{k} \left[\frac{(n+k)!}{n!k!}\right]^2 (\tanh\eta)^{4k} ,
\end{eqnarray}
which is less than one.  This is  due to the fact that we are not observing
the $y$ variable.  Our knowledge is less than complete.

\par
The standard way to measure this incompleteness is to calculate the
entropy defined as~\cite{neu32,fano57,wiya63}
\begin{equation}
          S = - Tr\left(\rho(x,r)  \ln[\rho(x,r)]\right)  ,
\end{equation}
which leads to
\begin{eqnarray}
&{}&    S = 2(n + 1)[(\cosh\eta)^2 \ln(\cosh\eta) -
                           (\sinh\eta)^2 \ln(\sinh\eta)] \nonumber \\[1ex]
  &{}& \hspace{10mm} - \left(\frac{1}{\cosh\eta}\right)^{2(n + 1)}
  \sum_{k} \frac{(n+k)!}{n!k!}\ln\left[\frac{(n+k)!}{n!k!}\right]
               (\tanh\eta)^{2k} .
\end{eqnarray}

Let us go back to the wave function given in Eq.(\ref{wf15}).  As is
illustrated in Fig.~\ref{restof66}, its localization property is dictated by
its Gaussian factor which corresponds to the ground-state wave function.  For
this reason, we expect that much of the behavior of the density matrix or
the entropy for the $n^{th}$ excited state will be the same as that for the
ground state with n = 0.  For this state, the density matrix is
\begin{equation}
  \rho_{\eta}(x,r) = \left(\frac{1}{\pi \cosh(2\eta)}\right)^{1/2}
  \exp{\left\{-\frac{1}{4}\left[\frac{(x + r)^2}{\cosh(2\eta)}
                    + (x - r)^2\cosh(2\eta)\right]\right\}} ,
\end{equation}
and the entropy is
\begin{equation}\label{entro11}
  S_{\eta} = 2\left[ (\cosh\eta)^2 \ln(\cosh\eta) - (\sinh\eta)^2 \ln(\sinh\eta)\right]  .
\end{equation}
The density distribution $\rho_{\eta}(x,x)$ becomes
\begin{equation}\label{eq21}
  \rho_{\eta}(x,x) = \left(\frac{1}{\pi \cosh(2\eta)}\right)^{1/2}
  \exp{\left(\frac{-x^2}{\cosh(2\eta)}\right) }.
\end{equation}
The width of the distribution becomes $\sqrt{\cosh(2\eta)}$, and the
distribution becomes wide-spread as $\eta$ becomes larger.
Likewise, the momentum distribution becomes wide-spread as can be seen in
Fig.~\ref{ctran33}.  This simultaneous increase in the momentum and position
distribution widths is due to our inability to measure the $y$ variable hidden
in Feynman's rest of the universe~\cite{fey72}.

\par
In their paper of 1990~\cite{kiwi90pl}, Kim and Wigner used the $x$ and $y$ variables
as the longitudinal and time-like variables respectively in the Lorentz-covariant
world. In the quantum world, it is a widely accepted view that there are no time-like
excitations.  Thus, it is fully justified to restrict the $y$ component to its
ground state as we did in Sec.~\ref{excited}.

\section{Space-time Entanglement}\label{spt}

The series given in Eq.(\ref{seri00}) plays the central role in the
concept of the Gaussian or continuous-variable entanglement, where
the measurement on one variable affects the quantum mechanics of
the other variable.  If one of the variables is not observed, it
belongs to Feynman's rest of the universe.
\par
\begin{figure}
\centerline{\includegraphics[scale=2.8]{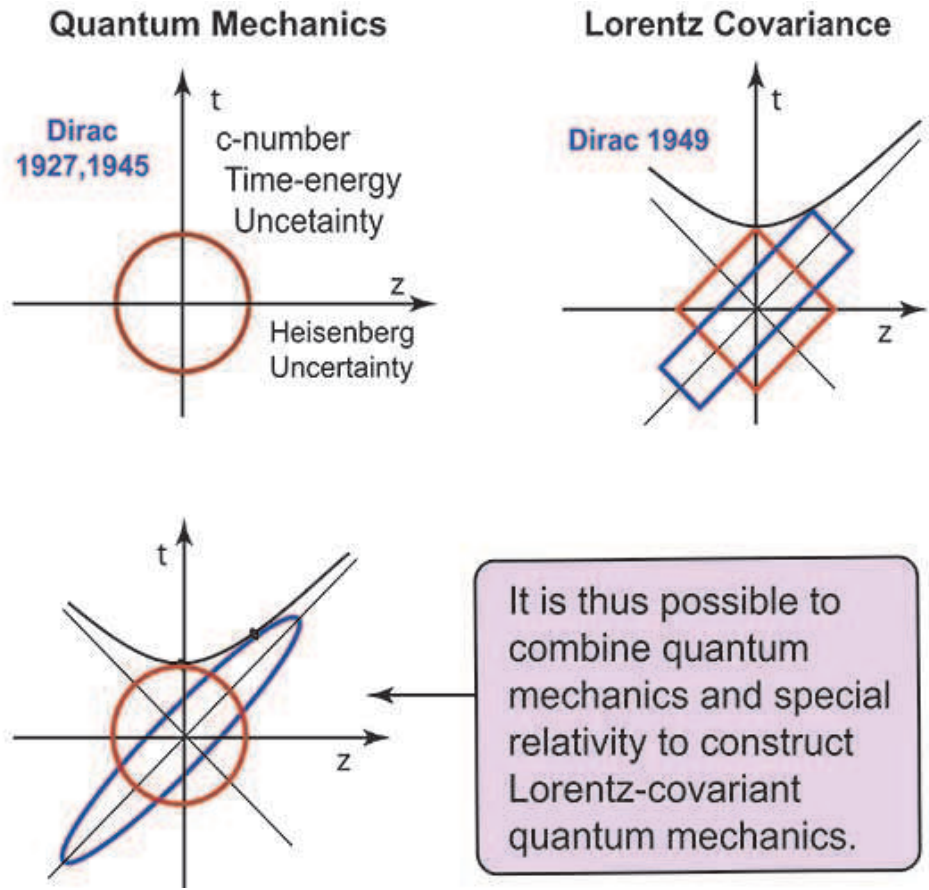}}
\caption{Dirac's form of Lorentz-covariant quantum mechanics.   In
addition to Heisenberg's uncertainty relation which allows excitations
along the spatial direction, there is the ``c-number'' time-energy
uncertainty without excitations.  This form of quantum mechanics
can be combined with Dirac's light-cone picture of Lorentz boost,
resulting in the Lorentz-covariant picture of quantum mechanics.
The elliptic squeeze shown in this figure can be called
the space-time entanglement.}\label{diracqm}
\end{figure}

The series of the form of Eq.(\ref{seri00}) was developed earlier for
studying harmonic oscillators in moving
frames~\cite{kn73,knp86,kno79jmp,kno79ajp,kiwi90pl,kn11symm}.
Here $z$ and $t$  are the space-like and time-like separations between
the two constituent particles bound together by a harmonic oscillator
potential. There are excitations along the longitudinal direction.   However,
no excitations are allowed along the time-like direction.  Dirac described this
as ``c-number'' time-energy uncertainty relation~\cite{dir27}.  Dirac in 1927
was talking about the system without special relativity.  In 1945~\cite{dir45},
Dirac attempted to construct space-time wave functions using harmonic oscillators.
In 1949~\cite{dir49}, Dirac introduced his light-cone coordinate system for
Lorentz boosts, telling that the boost is a squeeze transformation.
It is now possible to combine Dirac's three observations to construct the
Lorentz covariant picture of  quantum bound states, as illustrated in
Fig.~\ref{diracqm}.

\par
If the system is at rest, we use the wave function
\begin{equation}\label{wf81}
\psi_{0}^{n}(z,t) = \chi_{n}(z) \chi_{0}(t),
\end{equation}
which allows excitations along the $z$ axis, but no exciations along the
$t$ axis, according to Dirac's c-number time-energy uncertainty relation.

\par
If the system is boosted, the $z $ and $t$ variables are replaced by
$z'$ and $t'$ where
\begin{equation}
z'  = (\cosh\eta) z - (\sinh\eta)  t , \quad\mbox{and}\quad
t'  = -(\sinh\eta) z + (\cosh\eta) t .
\end{equation}
This is a squeeze transformation as in the case of Eq.(\ref{trans22}).
In terms of these space-time variables, the wave function of Eq.(\ref{wf15}),
can be written as
\begin{equation}\label{wf82}
\psi_{\eta}^{n}(z,t) = \chi_{n}(z') \chi_{0}(t'),
\end{equation}
and the series of Eq.(\ref{seri55}) then becomes
\begin{equation}\label{seri66}
\psi_{\eta}^n(z,t) = \left(\frac{1}{\cosh\eta}\right)^{(n+1)}
 \sum_{k} \left[\frac{(n+k)!}{n!k!}\right]^{1/2}
       (\tanh\eta)^{k} \chi_{k+n}(z)\chi_{k}(t) .
\end{equation}

Since the Lorentz-covariant oscillator formalism shares the same set of
formulas with the Gaussian entangled states, it is possible to explain
some aspects of space-time physics using the concepts and terminologies
developed in quantum optics as illustrated in Fig.~\ref{resonance}.

\par

The time-separation variable is a case in point.  The Bohr radius is a
well-defined spatial separation between the proton and electron in the hydrogen
atom.  However, if the atom is boosted, this radius picks up its time-like
separation.  This time-separation variable does not exist in the Schr\"odinger
picture of quantum mechanics.  However, this variable plays the pivotal role
in the covariant harmonic oscillator formalism.  It is gratifying to note
that this ``hidden or forgotten'' variable plays its role in the real world
while being entangled with the observable longitudinal variable.
With this point in mind, let us study some of the consequences of this
space-time entanglement.

\par

First of all, does the wave function of Eq.(\ref{wf82}) carry a probability
interpretation in the Lorentz-covariant world?  Since $dz dt = dz' dt'$,
the normalization
\begin{equation}\label{norm11}
\int |\psi_{\eta}^{n} (z,t)|^2 dt dz = 1 .
\end{equation}
This is a Lorentz-invariant normalization.   If the system is at rest,
the $z$ and $t$ variables are completely dis-entangled, and the spatial
component of the wave function satisfies  the Shcr\"odinger equation without
the time-separation variable.
\par
However, in the Lorentz-covariant world, we have to consider the inner
product
\begin{equation}\label{norm22}
  \left(\psi_{\eta}^{n} (z,t), \psi_{\eta'}^{m}(z,t)\right)
 = \int  \left[\psi_{\eta}^{n} (z,t)\right]^{*} \psi_{\eta'}^{m}(z,t) dz dt .
\end{equation}
The evaulation of this integral was carried out by Michael Ruiz in
1974~\cite{ruiz74}, and the result was
\begin{equation}
\left(\frac{1}{|\cosh(\eta - \eta')|}\right)^{n + 1} \delta_{nm} .
\end{equation}

In order to see the physical implications of this result, let us assume that
one of the oscillators is at rest with $\eta'= 0$ and the other is moving
with the velocity $\beta = \tanh(\eta)$.   Then the result is
\begin{equation}
  \left(\psi_{\eta}^n (z,t), \psi_{0}^m(z,t)\right) =
  \left(\sqrt{1 - \beta^2}\right)^{n + 1}  \delta_{nm} .
\end{equation}
Indeed, the wave functions are orthnormal if they are in the same Lorentz
frame.  If one of them is boosted, the inner product shows the effect of
Lorentz contraction.  We are familiar with the contraction $\sqrt{1 - \beta^2}$
for the rigid rod.   The ground state of the oscillator wave function is
contracted like a rigid rod.
\par
\begin{figure}
\centerline{\includegraphics[scale=2.4]{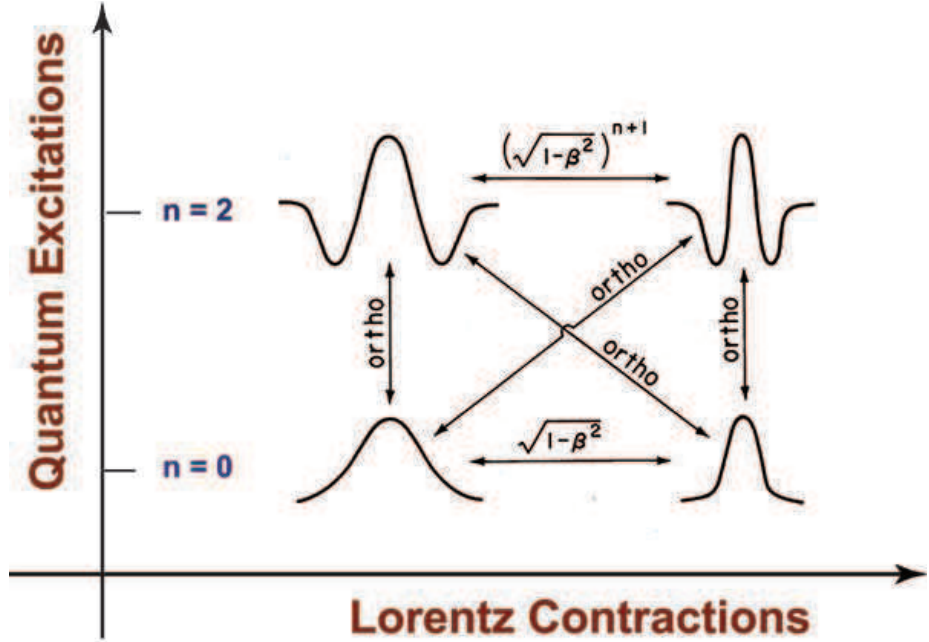}}
\caption{Orthogonality relations for two covariant oscillator wave
funtions.  The orthogonality relation is preserved for different frames.
However, they show the Lorentz contraction effect for two different
frames.}\label{ortho}
\end{figure}

The probability density $|\psi_{\eta}^{0}(z)|^2$ is for the oscillator in the
ground sate, and it has one hump.  For the $n^{th}$ excited state, there are
$(n + 1)$ humps.  If each hump is contracted like $\sqrt{1 - \beta^2}$, the
net contraction factor is $\left( \sqrt{1 - \beta^2} \right)^{n + 1}$ for the
$n^{th}$ excited state.  This result is illustrated in Fig.~\ref{ortho}.
\par

\par
With this understanding, let us go back to the entanglement problem.  The
ground state wave function takes the Gaussian form given in Eq.(\ref{gau00})
\begin{equation}\label{gau30}
 \psi_{0}(z,t) = \frac{1}{\sqrt{\pi}}
              \exp{\left\{-\frac{1}{2}\left(z^2 + t^2\right)\right\}} ,
\end{equation}
where the $x$ and $y$ variables are replaced by $z$ and $t$ respectively.
If Lorentz-boosted, this Gaussian function becomes squeezed
to~\cite{kn73,knp86,kno79jmp}
\begin{equation}\label{gau35}
 \psi_{\eta}^{0}(z,t) = \frac{1}{\sqrt{\pi}}
              \exp{\left\{-\frac{1}{4}\left[e^{-2\eta}(z + t)^2 +
               e^{2\eta}(z - t)^2\right]\right\}},
\end{equation}
leading to the series
\begin{equation} \label{seri30}
\frac{1}{\cosh\eta}\sum_{k} (\tanh{\eta})^k \chi_{k}(z) \chi_{k}(t) .
\end{equation}
According to this formula, the $z$ and $t$ variables are entangled in the same
way as the $x$ and $y$ variables are entangled.
\par
Here the $z$ and $t$ variables are space and time separations between two
particles bound together by the oscillator force.   The concept of the space
separation is well defined, as in the case of the Bohr radius.  On the
other hand, the time separation is still hidden or forgotten in the present
form of quantum mechanics.  In the Lorentz-covariant world, this variable
affects what we observe in the real world by entangling itself with the
longitudinal spatial separation.

 \par
In Chapter 16 of their book~\cite{walls08}, Walls and Milburn wrote down the
series of Eq.(\ref{seri00}) and discussed what would happen when the $\eta$
parameter becomes infinitely large.  We note that the series given in
Eq.(\ref{seri30}) shares the same expression as the form given by
Walls and Milburn, as well as other papers dealing with the Gaussian entanglement.
As in the case of Wall and Milburn, we are interested in what happens when
$\eta$ becomes very large.

\par

\begin{figure}
\centerline{\includegraphics[scale=4.8]{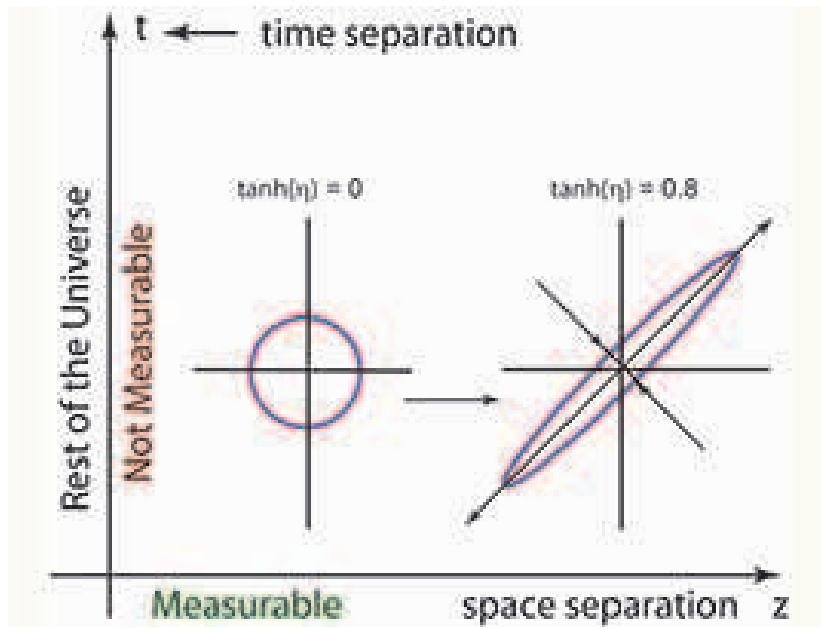}}
\caption{Feynman's rest of the universe.  This figure is the same as
that of Fig.~\ref{restof66}.  Here, the space variable $z$ and the time
variable $t$ are entangled.}\label{restof77}
\end{figure}

As we emphasized throughout the present paper, it is possible to study the
entanglement series using the squeezed Gaussian function given in Eq.(\ref{gau35}).
It is then possible to study this problem using the ellipse.  Indeed, we can
carry out the mathematics of entanglement using the ellipse shown Fig.~\ref{restof77}.
This figure is the same as that of Fig.~\ref{restof66},
but it tells about the entanglement of the space and time separations, instead
of the $x$ and $y$ variables.  If the particle is at rest with $\eta = 0$, the
Gaussian form corresponds to the circle in Fig.~\ref{restof77}.  When the
particle gains  speed, this Gaussian function becomes squeezed into an
ellipse.  This ellipse becomes concentrated along the light cone with $t = z$,
as $\eta$ becomes very large.

\par

\par

The point is that we are able to observe this effect in the real world.  These
days, the velocity of protons from high-energy accelators is very close to
that of light.  According to Gell-Mann~\cite{gell64}, the proton is a bound
state of three quarks.   Since quarks are confined in the proton, they have
never been observed, and the binding force must be like that of the harmonic
oscillator.  Furthermore, the observed mass spectra
of the hadrons exhbit the degeneracy of the three-dimentional harmonic
oscillator~\cite{fkr71}.  We use the word ``hadron'' for the bound state of
the quarks.  The simplest hadron is thus the bound state of two quarks.

\par

In 1969~\cite{fey69}, Feynman observed that the same proton, when moving
with its velocity close to that of light, can be regarded as a collection
of partons, with the following peculiar properties.

\begin{itemize}
\item[1.]  The parton picture is valid only for protons moving with velocity
           close to that of light.
\item[2.]  The interaction time between the quarks become dilated, and partons
           are like free particles.
\item[3.]   The momentum distribution becomes wide-spread  as the proton moves
            faster.  Its width is proportional to the proton momentum.
\item[4.]  The number of partons is not conserved, while the proton starts
            with a finite number of  quarks.
\end{itemize}

\par

\begin{figure}
\centerline{\includegraphics[scale=3.0]{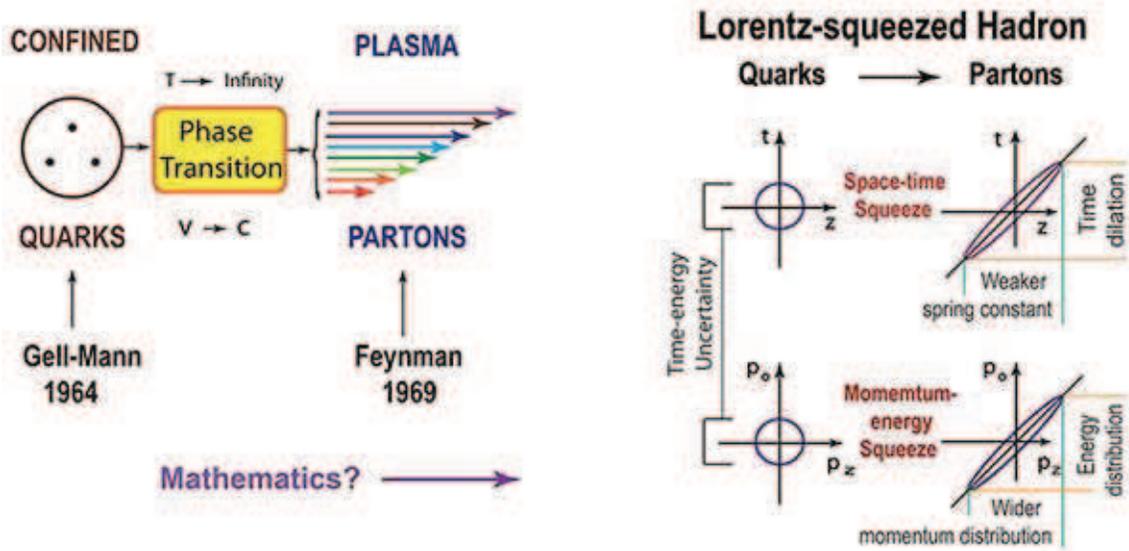}}
\caption{The transition from the quark to the parton model through
space-time entanglement.  When $\eta = 0$, the system is called the
quark model where the space separation  and the time separation are
dis-entangled.  Their entanglement becomes maximum when $\eta = \infty.$
The quark model is transformed continuously to the parton model as the
$\eta$ parameter increases from $0$ to $\infty$.  The mathematics of this
transformation is given in terms of circles and ellipses.}\label{quapar}
\end{figure}

\begin{figure}
\centerline{\includegraphics[scale=0.9]{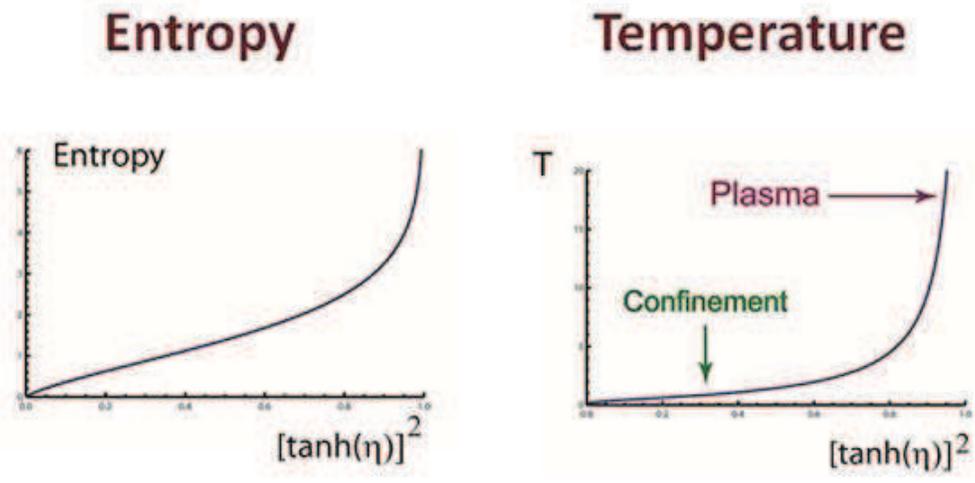}}
\caption{Entropy and temperature as functions of $[\tanh(\eta)]^2 = \beta^2$.
They are both zero when the hadron is at rest, but they become infinitely
large when the hadronic speed becomes close to that of light.  The curvature
for the temperature plot changes suddenly around $[\tanh(\eta)]^2 = 0.8$
indicating a phase transition.}\label{entemp}
\end{figure}

Indeed, Fig.~\ref{quapar} tells why the quark and parton models are two
limiting cases of one Lorentz-covariant entity.  In the oscillator regime,
the three-particle system can be reduced to two independent two-particle
systems~\cite{fkr71}. Also in the oscillator regime,  the momentum-energy
wave function takes the same form as the space-time wave function, thus with
the same squeeze or entanglement property as illustrated in this figure.
This leads to the wide-spread momentum distribution~\cite{knp86,kn77,kim89}.

\par
Also in Fig.~\ref{quapar}, the time-separation between the quarks becomes
large as $\eta$ becomes large, leading to a weaker spring constant.
This is why the partons behave like free particles~\cite{knp86,kn77,kim89}.

\par
As $\eta$ becomes very large, all the particles are confined into a narrow
strip around the light cone.  The number of particles is not constant for massless
particles as in the case of black-body radiation~\cite{knp86,kn77,kim89}.
\par
Indeed, the oscillator model explains the basic features of the hadronic
spectra~\cite{fkr71}.  Does the oscillator model tell the basic feature
of the parton distribution observed in high-energy laboratories?
The answer is YES.  In his 1982 paper~\cite{hussar81}, Paul
Hussar compared the parton distribution observed in high-energy laboratory
with the Lorentz-boosted Gaussian distribution.  They are close enough
to justify that the quark and parton models are two limiting cases of one
Lorentz-covariant entity.

\par

To summarize, the proton makes a phase transition from the bound state into a
plasma state as it moves faster, as illustrated in Fig.~\ref{quapar}.  The
un-observed time-separation variable becomes more prominent as $\eta$ becomes
larger.  We can now go back to the form of this entropy given in Eq.(\ref{entro11})
and calculate it numerically.  It is plotted against $(\tanh\eta)^2 = \beta^2$
in Fig.~\ref{entemp}. The entropy is zero when the hadron is at rest, and it
becomes infinite as the hadronic speed reaches the speed of light.

\par
Let us go back to the expression given in Eq.(\ref{eq14}).  For this ground
state, the density matrix becomes
\begin{equation}
\rho_{\eta} (z,z') =
    \left( \frac{1}{\cosh\eta}\right)^2\sum_{k}
       (\tanh\eta)^{2k} \chi_{k}(z)\chi_{k}(z') .
\end{equation}
We can now compare this expression with the density matrix for thermally
excited oscillator state~\cite{fey72}:
\begin{equation}
\rho_{\eta}(z,z') =
     \left(1 - e^{-1/T}\right)\sum_{k}\left[e^{-1/T}\right]^{k}
           \chi_{k}(z)\chi_{k}(z') .
\end{equation}
By comparing these two expressions, we arrive at
\begin{equation}
     [\tanh(\eta)]^2 = e^{-1/T} ,
\end{equation}
and thus
\begin{equation}
T = \frac{-1}{\ln\left[(\tanh\eta)^2\right]} .
\end{equation}
This temperature is also plotted against $(\tanh\eta)^2$ in Fig.~\ref{entemp}.
The temperature is zero if the hadron is at rest, but it becomes infinite
when the hadronic speed becomes close to that of light. The slope of the
curvature changes suddenly around $(\tanh\eta)^2 = 0.8$, indicating a phase
transition from the bound state to the plasma state.

\par

In this section, we have shown how useful the concept of entanglement is
in understanding the role of the time-separation in high energy hadronic
physics including Gell-Mann's quark model and Feynman's parton model as
two-limiting cases of one Lorentz-covariant entity.

\section*{Concluding Remarks}

The main point of this paper is the mathematical identity
\begin{equation}\label{conc11}
\frac{1}{\sqrt{\pi}}
              \exp{\left\{-\frac{1}{4}\left[e^{-2\eta}(x + y)^2 +
               e^{2\eta}(x - y)^{2}\right]\right\}} =
\frac{1}{\cosh\eta}\sum_{k} (\tanh{\eta})^k \chi_{k}(x) \chi_{k}(y) ,
\end{equation}
which says that the series of Eq.(\ref{seri00}) is an expansion of the Gaussian
form given in Eq.(\ref{i01}).

\par
The first derivation of this series was published in 1979~\cite{kno79ajp}
as a formula  from the Lorentz group.  Since this identity is not well
known, we explained in Sec.~\ref{excited} how this formula can be derived
from the generating function of the Hermite polynomials.
 \par
While the series serves useful purposes in understanding the physics of
entanglement, the Gaussian form can be used to transfer this idea to
high-energy hadronic physics.  The hadron, such as the proton, is a quantum
bound state.  As was pointed in Sec.~\ref{spt}, the squeezed  Gaussian
function of Eq.(\ref{conc11}) plays the pivotal role for hadrons moving with
relativistic speeds.

\par
The Bohr radius is a very important quantity in physics.  It is a spatial
separation between the proton and electron in the the hydrogen atom.  Likewise,
there is a space-like separation between constituent particles in a bound state
at rest.  When the bound state moves, it picks up a time-like component.
However, in the present form of quantum mechanics, this time-like separation
is not recognized.  Indeed, this variable is hidden in Feynman's rest of the
universe.   When the system is Lorentz-boosted, this variable entangles
itself with the measurable longitudinal variable.   Our failure to measure
this entangled variable appears in the form of entropy and temperature in the
real world.

\par
While harmonic oscillators are applicable to many aspects of quantum mechanics,
Paul A. M. Dirac observed in 1963~\cite{dir63} that the system of two oscillators
contains also the symmetries of the Lorentz group.  We discussed in this paper
one concrete case of Dirac's symmetry.  There are different languages for
harmonic oscillators, such as  the Schr\"odinger wave function, step-up
and step-down operators, the Wigner phase-space distribution function.  In
this paper, we used extensively a pictorial language with circles and ellipses.
\par
Let us go back to  Eq.(\ref{conc11}), this mathematical identity was published
in 1979 as a textbook material in the American Journal of Physics~\cite{kno79ajp},
and the same formula was later included in a textbook on the Lorentz
group~\cite{knp86}.  It is gratifying to note that the same formula serves
as a useful tool for the current literature in quantum information
theory~\cite{leonh10,furu11}.

\end{document}